
\input amstex
\documentstyle{amsppt}

\document
\magnification=\magstep1
\baselineskip=12pt
\let\<=\langle
\let\>=\rangle
\global\def\ssectitle#1\par{\bigbreak\medskip
  \leftline{\typc #1}
  \nobreak\bigskip\vskip-\parskip
  \message{#1}
  \noindent}
\def\today{\ifcase\month\or
  January\or February\or March\or April\or May\or June\or
  July\or August\or September\or October\or November\or December\fi
  \space\number\day, \number\year}

\def\AA{\Cal A}

\font\typc=cmbx10 scaled \magstep1  
\font\typg=cmcsc10 scaled \magstep2 
\font\typy=cmr10  
\font\type=cmbx10 scaled \magstep3  

\hrule
\medskip
\rightline{\sevenrm \today}
\medskip
\hrule
\bigskip
\bigskip
\bigskip

\centerline{\typg Sergey Fomin}

\medskip

{\typy
\settabs 2 \columns
\+  \quad\ \qquad Department of Mathematics,
& \quad\ \qquad \quad Theory of Algorithms Laboratory
\cr
\+ \quad\  Massachusetts Institute of Technology
& \quad\ \qquad \quad \quad \quad \quad
SPIIRAN, Russia \cr
}

\bigskip
\bigskip

\centerline{\typg Anatol N. Kirillov}

\medskip

{\typy
\settabs 2 \columns
\+  \qquad\ \qquad \qquad L.P.T.H.E.

& \quad\qquad \qquad  Steklov Mathematical Institute
\cr
\+ \qquad\ \qquad  Universit\'e Paris VI
& \quad\quad \qquad \quad \quad \quad
St.~Petersburg, Russia \cr
}

\bigskip

\topmatter
\bigskip
\bigskip
\centerline{\type{Universal exponential solution}}
\medskip
\centerline{\type{of the Yang-Baxter equation}}
\bigskip
\bigskip
\keywords Yang-Baxter equation, Schubert polynomials, symmetric functions
\endkeywords
\subjclass 05E
\endsubjclass
\endtopmatter

\ssectitle{Abstract}

Exponential solutions of the Yang-Baxter equation
give rise to generalized Schubert polynomials and corresponding
symmetric functions. We provide several descriptions
of the local stationary algebra defined by this equation.
This allows to construct various exponential solutions of
the YBE.
The $B_n$ and $G_2$ cases are also treated.

\ssectitle{1. Introduction}

Let $K$ be a field of zero chatacteristic.
Let $\AA = K[u_1,u_2,\dots](x,y,\dots)$ be the associative
algebra of formal power series in commuting variables $x$, $y$, \dots
with coefficients in a local stationary algebra \cite{V2}
with generators $u_1$, $u_2$,\dots. To be precise, we assume exactly
the following: non-adjacent generators $u_i$ and $u_j$ commute
whereas adjacent generators $u_i$ and $u_{i+1}$ are subject to
certain relations (perhaps infinitely many) which are invariant in $i$.

It was shown in \cite{FK1} (see also \cite{FS})
that a theory of generalized Schubert
polynomials and corresponding Stanley's symmetric functions can be developed
whenever one has a solution of the Yang-Baxter equation
$$ h_i(x)h_{i+1}(x+y)h_i(y) = h_{i+1}(y)h_i(x+y)h_{i+1}(x) $$
given by $h_i(x)=e^{xu_i}$.
In other words, this theory requires the condition
$$ e^{xa} e^{(x+y)b} e^{ya} = e^{yb} e^{(x+y)a} e^{xb} \tag{1} $$
to be satisfied by any pair of adjacent generators $a$ and $b$.

In this paper, a ``minimal'' set
of relations which would guarantee (1) is given.
(In other words, we characterize
the local stationary algebra $\AA_0$ defined by (1).)
This enables us to construct exponential solutions
of the YBE related to the following quotient algebras of $\AA_0$
(cf. \cite{FS, FK1, R}):

\noindent (i) the nilCoxeter algebra of the symmetric group;

\noindent (ii) the degenerate Hecke algebra $\Cal H_\infty(0)$;

\noindent (iii) the universal enveloping algebra of $U_+(gl(n))$;

\noindent (iv) the local Heisenberg algebra.

In Section 5, the relation between the YBE
and [generalized] Verma identities is treated.
Sections 4 and 6 contain some parallel results for the
$B_n$ and $G_2$ cases.

{\smc Acknowledgement.}
The authors are grateful to Nantel Bergeron, Richard Stanley and
Volkmar Welker for helpful discussions.
This work was completed when the authors were visiting LaBRI
at Universit\'e Bordeaux~I, France.

\ssectitle{2. Main results}

The equation (1) is actually an infinite set of conditions
on $a$ and $b$, given by taking coefficients of $x^ny^m$ on both sides
of (1). For example, equating the coefficients of $x^2y$ gives
the first nontrivial condition
$$ b^2a+ab^2+2aba=ba^2+a^2b+2bab\ . $$
We are going to write a minimal list of relations that imply (1).
This can be done in many different ways, as shown in Section 3.
For the purposes of this section, however, it suffices to have
the simplest of these characterizations.
To state it, we need some notation.

Let $C_0=a+b$ and  $C_{m+1}=[a,C_m]$ for $n=0,1,2,\dots$;
here $[\ ,\ ]$ stands for the commutator: $[f,g]=fg-gf$.
Thus
$$ C_1=[a,b]\ ,\ \ C_2=[a,[a,b]]\ ,\ \ C_3=[a,[a,[a,b]]]\ ,\ \dots\ $$

\proclaim{Theorem 1}
The condition {\rm (1)} is satisfied if and only if
for any odd $m$ the expression $C_m$ commutes with $a+b$.
\endproclaim

A proof of Theorem~1, along with other equivalent characterizations
of the algebra defined by (1), is given in Section 3.

This theorem enables us to construct various examples of exponential
solutions of the Yang-Baxter equation.

\proclaim{Example 1}\ \cite{FK1, R}\
Hecke algebras.\rm\ \
Suppose the generators $u_i$ of a certain albebra satisfy the relations
$u_i^2=\beta u_i$.
In our previous notation, it means that
$$a^2=\beta a\ \ ,\ \  b^2=\beta b\ . \tag{2}$$
Straightforward computations then give
$$
\align
& C_0=a+b , \\
& C_1=[a,b] , \\
& C_2=[a,[a,b]]=\beta(ab+ba)-2aba , \\
& C_3=[a,[a,[a,b]]]=\beta^2[a,b]=\beta^2 C_1 , \\
& C_4=\beta^2 C_2, \dots
\endalign
$$
Therefore, in order to satisfy (1),
one has to have, in addition to (2), the commutation
$$ [C_0,C_1]=0 \ . \tag{3}$$
Using (2) and its consequence $[a+b,ab+ba]=0$,
one can see that (3) is equivalent to the {\it Coxeter relation}
$aba=bab$.
In other words, an exponential solution of the Yang-Baxter equation
can be obtained from an algebra defined by
$$
\align
& u_iu_j=u_ju_i\ , |i-j|\geq 2\ ; \\
& u_i^2=\beta u_i \ ; \\
& u_iu_{i+1}u_i=u_{i+1}u_iu_{i+1}\ .
\tag{4}
\endalign
$$
\endproclaim

The main special cases are $\beta=0$ and $\beta=-1$.

\proclaim{Example 2}
The nilCoxeter algebra of the symmetric group.\rm\ \
This algebra (see \cite{FS}) can be defined as the algebra spanned by
permutations of $S_n$, with the multiplication rule
$$ w\cdot v = \cases \text{usual product\ \ $wv$\ \ if\ \ $l(w)+l(v)=l(wv)$} \\
		0\ ,\ \text{otherwise}
	\endcases
$$
where $l(w)$ is the length of a permutation $w$
(the number of inversions).
Another definition can be given in terms of generators $u_i$
(adjacent transpositions) satisfying the relations (4)
with $\beta=0$.
This example leads to the ordinary {\it Schubert polynomials}
of Lascoux and Sch\"utzenberger \cite{L} (see also \cite{M}),
as shown in \cite{FS} (cf. \cite{FK1}).
\endproclaim

\proclaim{Example 3}
The degenerate Hecke algebra.\rm\ \
Setting $\beta=-1$ gives a Hecke algebra $\Cal H_\infty(0)$.
The theory of generalized Schubert polynomials leads in this case
to the {\it Grothendieck polynomials}
of Lascoux and Sch\"utzenberger \cite{LS},
as shown in \cite{FK2}.
\endproclaim

\proclaim{Example 4} \ \cite{FK1}\
Universal enveloping algebra of $U_+(gl(n))$.\rm\ \
This algebra can be defined as the local algebra with generators
$u_1$, $u_2$, \dots
subject to Serre relations
$$ [u_i,[u_i,u_{i\pm 1}]]=0 \ .$$
Redenoting $a=u_i$ and $b=u_{i+1}$, we have
$$ [a,[a,b]]=0 \tag{5}$$
and
$$ [b,[b,a]]=0 \ .\tag{6}$$
This implies
$C_2=[a,[a,b]]=0$ and $[C_0,C_1]=[a+b,[a,b]]=0$
which guarantees the conditions of Theorem~1.
\endproclaim

\proclaim{Example 5} \
The local Heisenberg algebra.\rm\ \
This is a local stationary algebra with generators
$u_1$, $u_2$, \dots
satisfying
$$ [u_i,u_{i+1}]=\lambda_i $$
where $\lambda_i$ are some constants.
Since these relations imply both (5) and (6),
this is a quotient algebra of the algebra of the previous example.
\endproclaim

\ssectitle{3. Proof of the main theorem}

Define the sequence $\{T_n\}$ of elements of our algebra $\AA$ by
$$ T_0=1, \quad T_{n+1}=aT_n+T_nb \ .$$
Thus
$$ T_1=a+b\ ,\ \ T_2=a^2+2ab+b^2\ ,\ \ T_3=a^3+3a^2b+3ab^2+b^3\ ,\ \dots
$$

Let $C(x)$ and $T(x)$ be the following
generating functions for the sequences $\{C_n\}$ and $\{T_n\}$:
$$
\align
& C(x) = \sum_{n=0}^\infty \frac{C_n}{n!}x^n\ \ ,\\
& T(x) = \sum_{n=0}^\infty \frac{T_n}{n!}x^n\ \ .
\endalign
$$

\proclaim{Lemma 1}\quad
$ C(x) = e^{xa}(a+b)e^{-xa}\ ,\quad
T(x) = e^{xa}e^{xb}\ .
$
\endproclaim

{\bf Proof.}
The defining recurrences for $\{C_n(x)\}$ and $\{T_n(x)\}$
can be rewritten, in terms of generating functions, as
$$ C'(x)=aC(x)-C(x)a $$
and
$$ T'(x)=aT(x)+T(x)b\ .$$
Together with the constant terms $C(0)=a+b$ and $T(0)=1$ these
identities determine the functions $C(x)$ and $T(x)$ uniquely.
It only remains to check that the functions $C(x) = e^{xa}(a+b)e^{-xa}$
and $T(x) = e^{xa}e^{xb}$ do satisfy these identities.\ \qed

The following result is an extension of Theorem~1.

\proclaim{Theorem 2}
The following statements are equivalent:

\medskip

\item{\rm\quad {(i)}\qquad}
$e^{xa} e^{(x+y)b} e^{ya} = e^{yb} e^{(x+y)a} e^{xb}$\ ;

\medskip

\item{\rm\quad {(ii)}\qquad}
$[T(x),T(y)]=0$\ ;

\medskip

\item{\rm\quad {(iii)}\qquad}
$[C(x),C(y)]=0$\ ;

\medskip

\item{\rm\quad {(iv)}\qquad}
$[T(x),a+b]=0$\ ;

\medskip

\item{\rm\quad {(v)}\qquad}
$[C(x),a+b]=0$\ ;

\medskip

\item{\rm\quad {(vi)}\qquad}
$[T_n,T_m]=0$\ \ for all $n$ and $m$\ ;

\medskip

\item{\rm\quad {(vii)}\qquad}
$[C_n,C_m]=0$\ \ for all $n$ and $m$\ ;

\medskip

\item{\rm\quad {(viii)}\qquad}
$[T_n,a+b]=0$\ \ for all even $n$\ ;

\medskip

\item{\rm\quad {(ix)}\qquad}
$[C_m,a+b]=0$\ \ for all odd $m$\ ;

\medskip

\item{\rm\quad {(x)}\qquad}
$[T_n,T_{n+1}]=0$\ \ for all $n$\ ;

\medskip

\item{\rm\quad {(xi)}\qquad}
$[C_n,C_{n+1}]=0$\ \ for all $n$\ ;

\medskip

\item{\rm\quad {(xii)}\qquad}
$[a,[a,C_n]=[b,[b,C_n]]$\ \ for all even $n$\ ;

\medskip

\item{\rm\quad {(xiii)}\qquad}
$[\underbrace{a,[a,\ldots ,[a}_m,b]\ldots]]=
[\underbrace{b,[b,\ldots ,[b}_m,a]\ldots]]$\ \ for all even $m$\ .
\endproclaim

{\bf Proof.}

\medskip

\noindent $\boxed{\ \text{(i)}\ \Longleftrightarrow\ \text{(ii)}\ }$\quad
Restate (i) as $[e^{xa}e^{xb},e^{yb}e^{ya}]=0$
and note that $e^{xa}e^{xb}$ commutes with $e^{yb}e^{ya}$
iff it commutes with $(e^{yb}e^{ya})^{-1}=e^{-ya}e^{-yb}$
which means the same as commuting with $e^{ya}e^{yb}$.

\medskip

\noindent $\boxed{\ \text{(iii)}\ \Longleftrightarrow\ \text{(v)}\ }$\quad
Assume (v). Then
$$
\align
&e^{xa}(a+b)e^{-xa}e^{ya}(a+b)e^{-ya} \\
= &e^{ya}e^{(x-y)a}(a+b)e^{-(x-y)a}(a+b)e^{-ya} \\
= &e^{ya}(a+b)e^{(x-y)a}(a+b)e^{-(x-y)a}e^{-ya} \\
= &e^{ya}(a+b)e^{-ya}e^{xa}(a+b)e^{-xa}\ .
\endalign
$$
The implication $\text{(iii)}\Longrightarrow\text{(v)}$
is trivial since $a+b=C(0)$.

\medskip

\noindent $\boxed{\ \text{(ii)}\ \Longleftrightarrow\ \text{(vi)}\ }$\quad
Follows from Lemma.

\medskip

\noindent $\boxed{\ \text{(iii)}\ \Longleftrightarrow\ \text{(vii)}\ }$\quad
Follows from Lemma.

\medskip

\noindent $\boxed{\ \text{(ii)}\ \Longleftrightarrow\ \text{(iii)}\ }$\quad
The implication $\text{(ii)}\Longrightarrow\text{(iii)}$
follows from the identity
$$ C(x)=T'(x)(T(x))^{-1} \tag{7} $$
which is a formal consequence of Lemma:
$$ T'(x)(T(x))^{-1} = (e^{xa}ae^{xb}+e^{xa}be^{xb})(e^{-xb}e^{-xa})
= e^{xa}(a+b)e^{-xa} = C(x)\ .
$$
To prove the converse, rewrite (7) as
$$ T'(x)=C(x)T(x)\ ; $$
this means that $T_{n+1}$ is a certain noncommutative polynomial
in $C_0$,\dots,$C_n$,$T_1$,\dots,$T_n$; also $T_1=C_0$.
Therefore $T_{n+1}$ is a polynomial in $C_0$,\dots,$C_n$;
and if $C_i$'s commute, then so do $T_i$'s.

\medskip

\noindent $\boxed{\ \text{(ii)}\ \Longleftrightarrow\ \text{(iv)}\ }$\quad
Since $a+b=T(0)$, $\text{(ii)}\Longrightarrow\text{(iv)}$ is trivial.
Assume (iv); then (7) implies (v) implies (iii) implies (ii).

\medskip

\noindent $\boxed{\ \text{(vii)}\ \Longleftrightarrow\ \text{(xi)}\ }$\quad
The part $\text{(vii)}\Longrightarrow\text{(xi)}$ is trivial;
the part $\text{(xi)}\Longrightarrow\text{(vii)}$ follows from
the identity
$$
\align
&[C_j,C_i] = [[a,C_{j-1}],C_i]
=  - [[C_{j-1},C_i],a] - [[C_i,a],C_{j-1}] \\
=  -&[[C_{j-1},C_i],a] + [C_{i+1},C_{j-1}]
\tag{8}
\endalign
$$
by induction on $j-i=1,2,3,\dots$.

\medskip

\noindent $\boxed{\ \text{(vi)}\ \Longleftrightarrow\ \text{(x)}\ }$\quad
This is proved analogously to
$\text{(vii)}\ \Longleftrightarrow\ \text{(xi)}$; use
$$
\align
  &[T_j,T_i] = [[a,T_{j-1}]+T_{j-1}T_1,T_i] \\
= &[[a,T_{j-1}],T_i]+[T_{j-1}T_1,T_i] \\
= &-[[T_{j-1},T_i],a]-[[T_i,a],T_{j-1}]+[T_{j-1}T_1,T_i] \\
= &-[[T_{j-1},T_i],a]-[T_iT_1-T_{i+1},T_{j-1}]+[T_{j-1}T_1,T_i]
\tag{9}
\endalign
$$
instead of (8).

\medskip

\noindent $\boxed{\ \text{(vii)}\ \Longleftrightarrow\ \text{(ix)}\ }$\quad
Again, $\text{(vii)}\ \Longrightarrow\ \text{(ix)}$ is obvious.
Assume (ix). We will prove $[C_j,C_i]=0$ by induction on $i+j$.
If $i+j$ is even (and, say, $i<j$),
then (8) and the induction assumption give  $[C_j,C_i]=-[C_{j-1},C_{i+1}]$;
then repeat this argument until the indices coincide.
If $i+j$ is odd (now let us take $i>j$), then
repeatedly apply (8) and the induction hypothesis to get
$$ [C_j,C_i]=-[C_{j-1},C_{i+1}]=[C_{j-2},C_{i+2}]=\dots=\pm[C_0,C_{i+j}]=0\ ;$$
we used (ix) in the last step.

\medskip

\noindent $\boxed{\ \text{(vi)}\ \Longleftrightarrow\ \text{(viii)}\ }$\quad
Same argument as in $\text{(vii)}\ \Longleftrightarrow\ \text{(ix)}$,
with (8) replaced by (9).

\medskip

\noindent $\boxed{\ \text{(vii)}\ \Longleftrightarrow\ \text{(xii)}\ }$\quad
Use an identity
$$ [a,[a,C_n]]-[b,[b,C_n]] = (a-b)[C_0,C_n]+[C_n,C_0](a-b)+[C_n,C_1] $$
(one only needs to know that $C_0=a+b$ and $C_1=[a,b]$ to check this)
to prove the $\Longrightarrow$ part.
The same identity, together with (8) and induction on $i+j$, proves
the $\Longleftarrow$ part.

\medskip

\noindent $\boxed{\ \text{(xii)}\ \Longleftrightarrow\ \text{(xiii)}\ }$\quad
Induction on $n$ and $m$ for the implications $\Longleftarrow$
and $\Longrightarrow$, respectively.

This completes the proof of Theorem~2.
\ \qed

Theorem~1 is just the statement
\ $\boxed{\ \text{(i)}\ \Longleftrightarrow\ \text{(ix)}\ }$~.

\medskip

{\bf Comments. 1.}
The above proofs of $\text{(ii)}\ \Longleftrightarrow\ \text{(iv)}$
and $\text{(iii)}\ \Longleftrightarrow\ \text{(v)}$
could in fact be omitted;
e.g., we gave an independent proof of of
$\text{(ix)}\ \Longleftrightarrow\ \text{(vii)}$
whereas obviously $\text{(v)}\ \Longrightarrow\ \text{(ix)}$
and $\text{(vii)}\ \Longleftrightarrow\ \text{(iii)}$.

{\bf 2.} In view of condition (vii) of Theorem~2,
it is natural to suggest that the graded associative
algebra $\Cal A_0$
defined by (1) (or by any of (i)-(xiii))
is isomorphic to the algebra formally generated by pairwise
commuting elements $C_0$, $C_1$, $C_2$, \dots and an element
$a$ satisfying the conditions $[a,C_i]=C_{i+1}$;
the ranks are defined by $rk(a)=1$, $rk(C_i)=i+1$.

The last statement has been recently proved by Nantel
Bergeron~\cite{B}. It follows immediately that the Hilbert series
of $\Cal A_0$ is
$$ (1-t)^{-2}(1-t^2)^{-1}(1-t^3)^{-1}(1-t^4)^{-1}\cdot\cdot\cdot \ ,$$
and a linear basis is given by
$\{a^i C_0^{i_0} C_1^{i_1}\cdot\cdot\cdot\}$.
N.Bergeron also constructed a linear basis of {\it words}
in $\Cal A_0$.

\ssectitle{4. The $B_n$ case}

Many of the previous results can be extended to the $B_n$ case
where the analogue of the Yang-Baxter equation is
$$ h_2(x-y)h_1(x)h_2(x+y)h_1(y) = h_1(y)h_2(x+y)h_1(x)h_2(x-y) $$
(for another adjacent pairs the $A_n$-YBE is kept).
If we are interested in exponential solutions
$h_1(x)=e^{xa}$, $h_2(x)=e^{xb}$, then the YBE becomes
$$ [e^{xb}e^{xa}e^{xb},e^{yb}e^{ya}e^{yb}] = 0 \ . \tag{10} $$
It can be shown \cite{FK3} that the theory of [generalized]
Schubert/Grothendieck/Stanley polynomials can be constructed whenever
one has a local stationary algebra whose first two generators
satisfy (10) and any other adjacent generators satisfy (1).
We will show how solutions of (10) can be constructed using the
approach of Sections 2-3.

\proclaim{Example 6} \rm
Let the generators $a$ and $b$ satisfy the relations
$$ [b,[b,[b,a]]]=0 \ , \tag{11} $$
$$ [a,[b,[b,a]]]=0 \ , \tag{12} $$
and
$$ [a,[a,[b,a]]]=0 \ . \tag{13} $$
Note that $[a,[b,[b,a]]]=-[b,[a,[a,b]]]$, so (12) can be restated as
$$ [b,[a,[a,b]]]=0 \ . \tag{12'} $$
\endproclaim

\proclaim{Theorem 3}\quad
{\rm (11)-(13)} imply {\rm (10)}.
\endproclaim

{\bf Proof.} We start with introducing
formal power series
$$ R(x)=e^{xb}e^{xa}e^{xb} $$
and
$$ L(x)=R'(x)(R(x))^{-1}\ . $$
Similarly to the implication $\text{(iii)}\Longrightarrow\text{(ii)}$
of Theorem~2, $R(x)$'s form a commuting family
(cf.~(10)) provided so do $L(x)$'s. We are going to prove now that
(11)-(13) imply $\{L(x)\}$ is a commuting family.

Since
$$
\align
L(x) &= R'(x)(R(x))^{-1} = R'(x)R(-x)\\
& = (be^{xb}e^{xa}e^{xb} + e^{xb}ae^{xa}e^{xb} + e^{xb}e^{xa}be^{xb})
e^{-xb}e^{-xa}e^{-xb} \\
& = b + e^{xb}ae^{-xb} + e^{xb}e^{xa}be^{-xa}e^{-xb}\ ,
\endalign
$$
then
$$
L'(x) = e^{xb}[b,a]e^{-xb} + [b,e^{xb}e^{xa}be^{-xa}e^{-xb}]
+ e^{xb}e^{xa}[a,b]e^{-xa}e^{-xb}\ .
$$
Further straightforward calculations make a repeated use
of the identity
$$ e^{xf}ge^{-xf} = g + x[f,g] + x^2[f,[f,g]]/2 + x^3[f,[f,[f,g]]]/6 + \dots\ ,
$$
along with (11)-(13):
$$
\align
L'(x)  = &e^{xb}[b,a]e^{-xb} + [b,e^{xb}(b+x[a,b]+x^2[a,[a,b]]/2)e^{-xb}] \\
	 & + e^{xb}([a,b]+x[a,[a,b]])e^{-xb} \\
       = & x[b,e^{xb}([a,b]+x[a,[a,b]]/2)e^{-xb}] + xe^{xb}[a,[a,b]]e^{-xb} \\
       = & x[b,[a,b]+x[b,[a,b]]+x[a,[a,b]]/2] + x[a,[a,b]] \\
       = & x[a+b,[a,b]] \ .
\endalign
$$
To prove that $L(x)$ is a commuting family, we only need now
$$ [L(0),L'(x)] = 0 $$
which reduces to
$$ [a+2b,[a+b,[a,b]]] = 0 \ . \tag{14}$$
The latter follows instantly from (11), (12), (13), and (12$'$).
This completes the proof of the theorem.\ \qed

It should be noted that, in fact, (14) is the first (i.e., the
lowest-degree) condition on $a$ and $b$ in the ``$B_n$-universal''
algebra $\Cal B_0$ defined by (10).
There are no conditions in degree 5.
We hope to give an exact list of defining relations for $\Cal B_0$
in another publication.

\proclaim{Example 7} The nilCoxeter algebra of the
hyperoctahedral group.\rm\ \ (Compare to Example~2.)
Assume $a^2=b^2=0$. Then (11) and (13) are guaranteed,
and (12) reduces to $abab-baba=0$. We conclude that (11)-(13)
are satisfied in the algebra defined by
$$
\align
& u_iu_j=u_ju_i\ , |i-j|\geq 2; \\
& u_i^2=0 \ ; \\
& u_iu_{i+1}u_i=u_{i+1}u_iu_{i+1}\ , i\geq 2\ ; \\
& u_1u_2u_1u_2=u_2u_1u_2u_1
\endalign
$$
which is the nilCoxeter algebra of the
hyperoctahedral group (the definition is similar to the
$A_n$ case).
\endproclaim

\proclaim{Example 8} Universal enveloping algebra of $U_+(so(2n+1))$.
\rm\ \ (Compare to Example~4.)
This algebra can be defined as the local algebra with generators
$u_1$, $u_2$, \dots
subject to Serre relations
$$
\align
& u_iu_j=u_ju_i\ , |i-j|\geq 2; \\
& [u_i,[u_i,u_{i+1}]]=0 \ , i\geq 2\ ; \\
& [u_i,[u_{i+1},u_{i+1}]]=0 \ , i\geq 2\ ; \\
\tag{15}
\endalign
$$
$$ [b,[b,a]]=0 \ ; \tag{16}$$
$$ [a,[a,[a,b]]]=0 \  \tag{17}$$
where $a=u_1$, $b=u_2$.
To show that (10) is satisfied in this case,
just note that (16)-(17) immediately imply (11)-(13).
\endproclaim

\proclaim{Example 9} Universal enveloping algebra of $U_+(sp(2n))$.
\rm\ \
This algebra is defined by (15) together with
$$
\align
&  [a,[a,b]]=0 \ ; \\
& [b,[b,[b,a]]]=0 \ . \tag{18}
\endalign
$$
The condition (10) is satisfied, analogously to Example~8.
\endproclaim

\ssectitle{5. Verma relations}

Let $\{u_i\}$ be the generators of the universal enveloping algebra
of Example~4 (one could also use Example~5 instead).
Let $\{t_i\}$ be arbitrary constants. Define
$$ e_i = ln(1+t_iu_i)\ ; \tag{19}$$
then, surprisingly, the $e_i$'s provide an exponential solution
of the YBE. In other words,
$$ e^{x\ln(1+ta)}e^{(x+y)\ln(1+sb)}e^{y\ln(1+ta)}
= e^{y\ln(1+sb)}e^{(x+y)\ln(1+ta)}e^{x\ln(1+sb)}
\tag{20}$$
where $a=u_i$, $b=u_{i+1}$, $t=t_i$, $s=t_{i+1}$.
(The locality condition obviously holds.)
It is not trivial at all that the last identity follows from
(5) and (6).
We will prove it by showing that it is true when $x=n$ and $y=m$
are arbitrary nonnegative integers, in which case
it converts into the following {\it generalized Verma identity}
(cf. \cite{V1}).

\proclaim{Lemma 2}
Conditions {\rm (5)-(6)} imply
$$ (1+ta)^n (1+sb)^{n+m} (1+ta)^m = (1+sb)^m (1+ta)^{n+m} (1+sb)^n \ .
\tag{21} $$
\endproclaim

Let us first make clear why (21) implies (20).
The equality (20) can be rewritten as
$$ \sum\Sb i,j,k,l\\ i+j\leq k+l\endSb P_{ijkl}(a,b) x^iy^jt^ks^l=0 $$
where $P_{ijkl}(a,b)$ are some {\it non-commutative polynomials}
in $a$ and $b$. We need $P_{ijkl}(a,b)=0$ for any $i$, $j$, $k$, and $l$.
Let us fix $k$ and $l$. Then
$$ \sum P_{ijkl}(a,b) x^iy^j = 0 \tag{22} $$
is a {\it finite} identity about $a$ and $b$; the finiteness is
ensured by the condition $i+j\leq k+l$.
The condition (21) means that (22) is true for a ny nonnegative integers
$x=n$ and $y=m$.
Since the matrix $\{n^im^j\}$ whose rows are indexed by pairs $(i,j)$
and columns by pairs $(n,m)$ has maximal rank, it follows that
all $P_{ijkl}(a,b)$ are 0.

Now it remains to prove Lemma 2.
First note that the elements $a'=1+ta$ and $b'=1+sb$
satisfy the same relations (5) and (6) as $a$ and $b$ do;
then it suffices to show the ordinary Verma identity
$$ a^nb^{n+m}a^m = b^ma^{n+m}b^n\ . \tag{23} $$

\proclaim{Lemma 3 {\rm(cf. \cite{S})}}
Assume that, in some associative algebra,
$ab-ba=\lambda$ where $\lambda$ commutes with both $a$ and $b$.
Then, for any nonnegative integer $n$,
$$ a^nb^n = (ba+\lambda)(ba+2\lambda)\cdot\cdot\cdot(ba+n\lambda) $$
and
$$ b^na^n = ba(ba-\lambda)\cdot\cdot\cdot(ba-(n-1)\lambda)\ .\ \qed $$
\endproclaim

This lemma immediately implies that $a^nb^n$ commutes with
$b^ma^m$ which is exactly (23).

In fact, the following generalization of (23) is true.

\proclaim{Lemma 4 \cite{F}}
Under assumptions of Lemma~3, any two ``balanced'' words
in $a$ and $b$ (that is, words containing as many $a$'s as $b$'s)
commute.\ \qed
\endproclaim

\proclaim{Conjecture}
The solution (19) is a universal exponential solution of the YBE.
In other words, the associative algebra generated by the elements $e_i$
is isomorphic to the local stationary algebra $\Cal A_0$
defined by the condition (1).
\endproclaim

The above results can be generalized to the $B_n$ case.
The $B_n$-Verma relations are
$$ a^n b^{2n+m} a^{n+m} b^m = b^m a^{n+m} b^{2n+m} a^n $$
where, as before, one can make substitutions $a\leftarrow 1+ta$
and $b\leftarrow 1+sb$.
Consequently, the elements $e_i=ln(1+t_iu_i)$
give an exponential solution of the $B_n$-YBE.

\ssectitle{6. The $G_2$ case}

Let us denote
$$\eqalignno{
&{\Cal P}(G_2)=\{~a,b~|~a^2=\beta a,~b^2=\beta b,~ababab=bababa~\},\cr
&U_+(G_2)=\{~a,b~|~[a,[a,[a,[a,b]]]]=0,~[b,[b,a]]=0~\}.}
$$
The following results can be verified.

{\it Exponential solutions of the Yang-Baxter equation.}
Let us define $h_1(x):=\exp (xa)$, $h_2(x):=\exp (xb)$,
where $a$ and $b$ are either the generators of the algebra ${\Cal P}(G_2)$
or those of $U_+(G_2)$.

Then
$$\eqalignno{
&h_1(x)h_2(3x+y)h_1(2x+y)h_2(3x+2y)h_1(x+y)h_2(y)\cr
&=h_2(y)h_1(x+y)h_2(3x+2y)h_1(2x+y)h_2(3x+y)h_1(x)\ .}
$$

{\it Generalized Verma relations.}
In the algebra $U_+(G_2)$,
$$
a^n b^{3n+m} a^{2n+m} b^{3n+2m} a^{n+m} b^m=
b^m a^{n+m} b^{3n+2m} a^{2n+m} b^{3n+m} a^n
\ .
$$
The same formula is true with $a$ and $b$ replaced by
$1+ta$ and $1+sb$, respectively.

\ssectitle{References}

\item{\quad {[B]}\quad}
N.Bergeron, private communication.

\item{\quad {[F]}\quad}
S.Fomin,
Duality of graded graphs,
Report No.15 (1991/92), {\it Institut Mittag-Leffler},
1992; submitted to {\it J.Algebr.Combinatorics}.

\item{\quad {[FK1]}\quad}
S.Fomin, A.N.Kirillov, The Yang-Baxter equation, symmetric functions,
and Schubert polynomials, to appear in {\it Proceedings of the 5th
International Conference on Formal Power Series and Algebraic Combinatorics},
Firenze, 1993.

\item{\quad {[FK2]}\quad}
S.Fomin, A.N.Kirillov, Grothendieck polynomials and the Yang-Baxter equation,
manuscript, 1993.

\item{\quad {[FK3]}\quad}
S.Fomin, A.N.Kirillov, Combinatorial $B_n$-analogues of Schubert polynomials,
manuscript, 1993.

\item{\quad {[FS]}\quad}
S.Fomin, R.P.Stanley, Schubert polynomials and the nilCoxeter algebra,
{\it Advances in Math.}, to appear; see also
{\it Report No.18 {\rm (1991/92),} Institut Mittag-Leffler},
1992.

\item{\quad {[L]}\quad}
A.Lascoux, Polyn\^omes de Schubert. Une approche historique,
{\it S\'eries formelles et combinatoire alg\'ebri\-que},
P.Leroux and C.Reutenauer, Ed., Montr\'eal, LACIM, UQAM, 1992, 283-296.

\item{\quad {[LS]}\quad}
A.Lascoux, M.P.Sch\"utzenberger,
D\'ecompositions dans l'alg\`ebre des differences divis\'ees,
preprint, 1989.

\item{\quad {[M]}\quad}
I. Macdonald, {\it Notes on Schubert polynomials},
Laboratoire de combinatoire et d'in\-for\-ma\-ti\-que math\'ema\-tique
(LACIM), Universit\'e du Qu\'ebec \`a Montr\'eal, Montr\'eal, 1991.

\item{\quad {[R]}\quad}
J.D.Rogawski,
On modules over the Hecke algebras of a $p$-adic group,
\it Inventiones Math. \bf 79 \rm (1985), 443.

\item{\quad {[S]}\quad}
R.P.Stanley,
On the number of reduced decompositions of elements of Coxeter groups,
\it European J. Combin. \bf 5 \rm (1984), 359-372.

\item{\quad {[V1]}\quad}
D. Verma,
Structure of certain induced representations of complex
semisimple Lie algebras, {\it Bull. Amer. Math. Soc.}
{\bf 74} (1968), 160-166.

\item{\quad {[V2]}\quad}
A.M.Vershik,
Local stationary algebras,
\it Amer. Math. Soc. Transl. \rm (2) \bf 148 \rm (1991), 1-13.

\enddocument